\documentstyle[11pt]{article}
\begin{document}

\title{\textbf{The generalized second law of gravitational
thermodynamics on the apparent horizon in $f(R)$-gravity}}

\author{K. Karami$^{1,2}$\thanks{KKarami@uok.ac.ir} ,
M.S. Khaledian$^{1}$\thanks{MS.Khaledian@uok.ac.ir} , N. Abdollahi$^{1}$\\\\
$^{1}$\small{Department of Physics, University of Kurdistan,
Pasdaran St., Sanandaj, Iran}\\$^{2}$\small{Research Institute for
Astronomy $\&$ Astrophysics of Maragha (RIAAM), Maragha, Iran}\\
}

\maketitle

\begin{abstract}
We investigate the generalized second law (GSL) of thermodynamics in
the framework of $f(R)$-gravity. We consider a FRW universe filled
only with ordinary matter enclosed by the dynamical apparent horizon
with the Hawking temperature. For a viable modified gravity model as
$f(R)=R-\alpha/R+\beta R^{2}$, we examine the validity of the GSL
during the early inflation and late acceleration eras. Our results
show that for the selected $f(R)$-gravity model minimally coupled
with matter, the GSL in the early inflation epoch is satisfied only
for the special range of the equation of state parameter of the
matter. But in the late acceleration regime, the GSL is always
respected.
\end{abstract}

\noindent\textbf{PACS numbers:} 04.50.Kd\\
\noindent\textbf{Keywords:} Modified theories of gravity

\clearpage
\section{Introduction}
In the last decade, the cosmological observations have confirmed the
existence of the early inflationary epoch and the accelerated
expansion of the present universe \cite{Riess}. The proposals that
have been put forth to explain these interesting discoveries can
basically be classified into two categories. One is to assume the
existence of an exotic energy with negative pressure, named dark
energy (DE). However, the problem of DE is one of the hardest and
unresolved problems in modern theoretical physics (see
\cite{Padmanabhan,Sahni} and references therein).

Another alternative is to modify Einstein's general relativity (GR)
theory, named modified gravity. One such modification is the $f(R)$
theory, where the Ricci scalar $R$ in the Einstein-Hilbert action is
generalized to an arbitrary function $f$ of $R$ (for a good review
see \cite{Sotiriou} and references therein). There are also some
other classes of modified gravities containing $f(\mathcal{G})$,
$f(R,\mathcal{G})$ and $f(T)$ which are considered as gravitational
alternatives for DE \cite{Capozziello,NojiriOdin,husawiski,noj
abdalla,Noj2,bengochea}. Here, ${\mathcal
G}=R_{\mu\nu\rho\sigma}R^{\mu\nu\rho\sigma}-4R_{\mu\nu}R^{\mu\nu}+R^2$
is the Gauss-Bonnet invariant term. Also $R_{\mu\nu\rho\sigma}$ and
$R_{\mu\nu}$ are the Riemann and Ricci tensors, respectively, and
$T$ is the torsion scalar. The modified gravity can explain
naturally the unification of earlier and later cosmological epochs
without resorting to the DE \cite{Sotiriou}. Moreover, modified
gravity may serve as dark matter (DM) \cite{Sobouti}.

Thermodynamical interpretation of gravity is one of the other
interesting issues in modern cosmology. It was shown that the
differential form of the Friedmann equation in the Einstein gravity
can be written in the form of the first law of thermodynamics
(Clausius relation) $-{\rm d}E=T_A{\rm d} S_A$ on the apparent
horizon $\tilde{r}_{\rm A}$. Here, $T_{\rm A}=1/(2\pi \tilde{r}_{\rm
A})$ and $S_A=\frac{A}{4G}$ are the Hawking temperature and entropy
on the apparent horizon, respectively, and $A$ is the area of the
horizon \cite{Cai05}. Also ${\rm d}E$ is the amount of the energy
flow through the fixed apparent horizon. Investigation on the deep
connection between gravity and thermodynamics has been also extended
to some modified gravity theories like the $f(\mathcal{G})$ theory
\cite{Cai05}, scalar-tensor gravity and $f(R)$-gravity
\cite{Akbar12}, Lovelock theory \cite{Akbar} and braneworld
scenarios (such as DGP, RSI and RSII) \cite{Sheykhi1}.

Note that in thermodynamics of apparent horizon in the Einstein
gravity, the geometric entropy is assumed to be proportional to its
horizon area, $S_A=\frac{A}{4G}$ \cite{Cai05}. However, this
definition for the other modified gravity theories is changed. For
instance, the geometric entropy in $f(R)$-gravity is given by
$S_A=\frac{Af_R}{4G}$ \cite{Wald}, where the subscript $R$ denotes a
derivative with respect to the curvature scalar R. In
$f(T)$-gravity, it was shown that when $f_{TT}$ is small, the first
law of black hole thermodynamics is satisfied approximatively and
the entropy of horizon is $S_A=\frac{Af_T}{4G}$ \cite{Miao}, where
subscript $T$ denotes a derivative with respect to the torsion
scalar $T$.

Besides the first law, the GSL of gravitational thermodynamics in
the accelerating universe driven by the DE or due to the modified
gravity has been studied extensively in
\cite{Izquierdo1}-\cite{Geng}. The GSL of thermodynamics like the
first law is a universal principle governing the universe.

Here, our aim is to investigate the GSL of thermodynamics in the
framework of $f(R)$-gravity for a Friedmann-Robertson-Walker (FRW)
universe filled with the ordinary matter. To do this, in section 2,
we briefly review the $f(R)$-gravity. In section 3, we investigate
the GSL of thermodynamics on the dynamical apparent horizon with the
Hawking temperature. In section 4, we examine the validity of the
GSL for a viable $f(R)$ model. Section 5 is devoted to conclusions.

\section{$f(R)$-gravity}
In $f(R)$-gravity, the modified Einstein-Hilbert action in the
Jordan frame is given by \cite{Sotiriou}
\begin{equation}\label{action}
    S_{\rm J}=\frac{1}{2k^2}\int {\rm d}^4x\sqrt{-g}~f(R)+S_{\rm m},
\end{equation}
where $k^2=8\pi G$. Also $g$ and $S_{\rm m}$ are the determinant of
metric $g_{\mu\nu}$ and the matter action, respectively.

Taking variation of the action (\ref{action}) with respect to
$g_{\mu\nu}$ leads to the corresponding field equations in
$f(R)$-gravity
\begin{equation}\label{evo}
 R_{\rm\mu\nu}f_{\rm R}-\frac{1}{2}g_{\rm\mu\nu}f-\nabla_{\rm\mu}\nabla_{\rm \nu}f_{\rm R}
 +g_{\rm\mu\nu}\nabla^{2}f_{\rm R}=k^2T_{\rm\mu\nu}^{(\rm m)},
\end{equation}
where $T^{\mu}_{\nu}={\rm diag}(-\rho_{\rm m},p_{\rm m},p_{\rm
m},p_{\rm m})$ is the matter energy-momentum tensor in the perfect
fluid form.

Now we consider a spatially non-flat universe described by the FRW
metric
\begin{equation}\label{FRW 1} {\rm d}s^2 = -{\rm
d}t^2+a^2(t)\left(\frac{{\rm d}r^2}{1-Kr^2} + {r}^2{\rm
d}\Omega^2\right),
\end{equation}
where $K=0,1,-1$ represent a flat, closed and open universe,
respectively. Substituting the FRW metric (\ref{FRW 1}) in the field
equations (\ref{evo}) yields the Friedmann equations in
$f(R)$-gravity as
\begin{eqnarray}
&&k^2\rho_{\rm m}=\frac{f}{2}-3\left( \dot{H}+H^{2}-H\frac{{\rm d}}{{\rm d} t}\right)f_{\rm R},\label{ro}\\
&&k^2p_{\rm
m}=-\frac{f}{2}+\left(\dot{H}+3H^{2}+\frac{2K}{a^{2}}-2H\frac{\rm
d}{{\rm d} t}-\frac{{\rm d}^{2}}{{\rm d} t^{2}}\right)f_{\rm
R},\label{p}
\end{eqnarray}
where
\begin{equation}
R=6\left(\dot{H}+2H^2+\frac{K}{a^2}\right),
\end{equation}
and ${\rm H}=\dot{a}/a$ is the Hubble parameter. Also the dot
denotes a derivative with respect to cosmic time $t$. The Friedmann
equations (\ref{ro}) and (\ref{p}) can be rewritten in the standard
form as \cite{Capozziello2}
\begin{equation}
{\textsl{H}}^2+\frac{K}{a^{2}}=\frac{k^2}{3}\rho_{\rm t},\label{Fr}
\end{equation}
\begin{equation}
\dot{H}-\frac{K}{a^{2}}=-\frac{k^2}{2}\big(\rho_{\rm t}+p_{\rm
t}\big).\label{Hdot}
\end{equation}
Here, $\rho_{\rm t}$ and $p_{\rm t}$ are the total energy density
and pressure defined as
\begin{eqnarray}
&&\rho_{\rm t}=\frac{\rho_{\rm m}}{f_{\rm R}}+\frac{\rho_{\rm R}}{k^2},\\
&&p_{\rm t}=\frac{p_{\rm m}}{f_{\rm R}}+\frac{p_{\rm R}}{k^2},
\end{eqnarray}
and $\rho_R$ and $p_R$ are the energy density and pressure due to
the curvature contribution defined as
\begin{equation}\label{roR}
\rho_{\rm R}= \frac{1}{f_{\rm R}}\left[-\frac{1}{2}(f-R f_{\rm
R})-3H\dot{f}_{\rm R}\right],
\end{equation}
\begin{equation}\label{pR}
p_{\rm R}=\frac{1}{f_{\rm R}}\left[\frac{1}{2}(f-R f_{\rm
R})+2H\dot{f}_{\rm R}+\ddot{f}_{\rm R}\right].
\end{equation}
Note that if $f(R)=R$, from Eqs. (\ref{roR}) and (\ref{pR}) we have
$\rho_{\rm R}=0$ and $p_{\rm R}=0$ then Eqs. (\ref{Fr}) and
(\ref{Hdot}) transform to the usual Friedmann equations in the
Einstein gravity.

The energy conservation laws are given by
\begin{equation}
\dot{\rho}_{\rm m}+3H\left( \rho_{\rm m}+p_{\rm m}\right)
=0,\label{ec}
\end{equation}
\begin{equation} \label{tot con}
\dot{\rho}_{\rm t}+3H\left( \rho_{\rm t}+p_{\rm t}\right) =0.
\end{equation}
Note that here the energy density and pressure due to the curvature
contribution satisfies the following energy equation
\begin{equation}\label{R con}
\dot{\rho}_{\rm R}+3H\left( \rho_{\rm R}+p_{\rm R}\right) =\frac{\dot{f}_{\rm R}}{f_{\rm R}^2}\rho_{\rm m}.
\end{equation}

\section{GSL in $f(R)$-gravity}
Here, we investigate the validity of the GSL in the framework of
$f(R)$-gravity. According to the GSL, entropy of the matter inside
the horizon beside the entropy associated with the surface of
horizon should not decrease during the time \cite{Cai05}. We
consider a FRW universe filled only with the ordinary matter. We
further assume the boundary of the universe to be enclosed by the
dynamical apparent horizon with the Hawking temperature.

The dynamical apparent horizon in FRW universe is given by
\cite{Poisson}
\begin{equation}\label{ra}
\tilde{r}_{\rm A}=\left( H^{2}+\frac{K}{a^{2}}\right)^{-1/2},
\end{equation}
which for the flat universe $K = 0$, it is same as the Hubble
horizon, i.e. $\tilde{r}_{\rm A}=H^{-1}$.

Following \cite{Cai05}, the associated Hawking temperature on the
apparent horizon is given by
\begin{equation}
T_{\rm A}=\frac{1}{2\pi \tilde{r}_{\rm
A}}\left(1-\frac{\dot{\tilde{r}}_{\rm A}}{2H\tilde{r}_{\rm
A}}\right),\label{ta}
\end{equation}
where $\frac{\dot{\tilde{r}}_{A}}{2H\tilde{r}_{A}}<1$ ensure that
the temperature is positive. Recently Cai et al. \cite{Cai09} proved
that the apparent horizon of the FRW universe has an associated
Hawking temperature (\ref{ta}). They also showed that this
temperature can be measured by an observer with the Kodoma vector
inside the apparent horizon.

The entropy of the matter inside the horizon is given by the Gibbs'
equation \cite{Izquierdo1}
\begin{equation}\label{gi}
T_{\rm A}{\rm d}S_{\rm m}={\rm d}E_{\rm m}+p_{\rm m}{\rm d}{V},
\end{equation}
where $E_{\rm m}=\rho_{\rm m}{V}$ and $V=\frac{4\pi}{3}
\tilde{r}_{\rm A}^{3}$ is the volume containing the matter with the
radius of the dynamical apparent horizon $\tilde{r}_{\rm A}$. Taking
time derivative of Eq. (\ref{gi}) and using the energy equation
(\ref{ec}) one can find
\begin{equation}\label{S mat f}
T_{\rm A}\dot{S_{\rm m}}=(\rho_{\rm m}+p_{\rm m})(\dot{V}-3HV).
\end{equation}
Substituting Eqs. (\ref{ro}) and (\ref{p}) into the above relation
gives
\begin{equation}
T_{\rm A}\dot{S}_{\rm m}=\frac{\tilde{r}_{\rm
A}^{2}}{2G}\Big(\dot{\tilde{r}}_{\rm A}-H\tilde{r}_{\rm A}\Big)
\left(\frac{2K}{a^{2}}-2\dot{H}+H\frac{\rm d}{{\rm d} t}-\frac{{\rm
d}^{2}}{{\rm d} t^{2}}\right)f_{\rm R}.\label{sma}
\end{equation}
The horizon entropy in $f(R)$-gravity is given by \cite{Wald}
\begin{equation}\label{S hor}
S_{\rm A}= \frac{Af_{\rm R}}{4G},
\end{equation}
where ${\rm A}=4\pi\tilde{r}_{\rm A}^{2}$ is the area of the
apparent horizon. Taking time derivative of Eq. (\ref{S hor}) and
using (\ref{ta}), one can get the evolution of the horizon entropy
as
\begin{equation}\label{saa}
T_{\rm A}\dot{S}_{\rm A}=\frac{1}{4GH}\left(2H\tilde{r}_{\rm
A}-\dot{\tilde{r}}_{\rm A}\right)\left(\frac{2\dot{\tilde{r}}_{\rm
A}}{\tilde{r}_{\rm A}}+\frac{\rm d}{{\rm d} t}\right)f_{\rm R}.
\end{equation}
Now we can calculate the GSL due to different contributions of the
matter and horizon. Adding Eqs. (\ref{sma}) and (\ref{saa}) and
using the auxiliary relations
\begin{eqnarray}
&&\dot{\tilde{r}}_{\rm A}=H\tilde{r}_{\rm A}^{3}\left(\frac{K}{a^{2}}-\dot{H}\right),\label{dra}\nonumber\\
&&H\tilde{r}_{\rm A}-\dot{\tilde{r}}_{\rm A}=H\tilde{r}_{\rm A}^3(\dot{H}+H^2),\label{dra1}\nonumber\\
&&2H\tilde{r}_{\rm A}-\dot{\tilde{r}}_{\rm A}=H\tilde{r}_{\rm
A}^3\left(\frac{K}{a^{2}}+\dot{H}+2H^2\right),\label{dra2}
\end{eqnarray}
one can get the GSL in $f(R)$-gravity as
\begin{eqnarray}\label{S Tot}
T_{\rm A}\dot{S}_{\rm
tot}=\frac{1}{4G}\left(H^{2}+\frac{K}{a^{2}}\right)^{-\frac{5}{2}}
\left\{2H\left(\frac{K}{a^{2}}-\dot{H}\right)^{2}f_{\rm R}\right.
~~~~~~~~~~~~~~~~~~~~~~~~~~~~~~~~~~~~~~~~~~~~~~~~~~\nonumber\\
\left.+\left[\frac{K}{a^2}\left(\frac{K}{a^2}+\dot{H}+3H^{2}\right)-\dot{H}H^{2}\right]\dot{f}_{\rm
R} +2H(\dot{H}+H^{2})\ddot{f}_{\rm R}\right\},
\end{eqnarray}
where $S_{\rm tot}=S_{\rm m}+S_{\rm A}$. Equation (\ref{S Tot})
shows that the validity of the GSL, i.e. $T_{\rm A}\dot{S}_{\rm
tot}\geq 0$, depends on the $f(R)$-gravity model. For instance, in
the Einstein gravity, i.e. $f(R)=R$, the GSL (\ref{S Tot}) yields
\begin{equation}\label{GSL-R}
T_{\rm A}\dot{S}_{\rm tot}=\frac{1}{G}\left[
\frac{H(\dot{H}-\frac{K}{a^{2}})^{2}}{2(H^{2}+\frac{K}{a^{2}})^{5/2}}\right]\geq
0,
\end{equation}
which shows that the GSL is always satisfied throughout the history
of the universe. Using Eqs. (\ref{Hdot}) and ({\ref{ra}), the above
relation can be rewritten as
\begin{equation}
T_{\rm A}\dot{S}_{\rm tot}=8\pi^{2}GH\tilde{r}_{\rm A}^{5}(
\rho_{\rm m}+p_{\rm m})^{2}\geq 0,
\end{equation}
which is same as that obtained in \cite{Karami1}.

\section{GSL for a viable $f(R)$ model}

Here, we would like to examine the validity of the GSL (\ref{S Tot})
for a viable $f(R)$ model given by
\begin{equation}
f(R)=R-\frac{\alpha}{R}+\beta R^{2},\label{fR}
\end{equation}
where $\alpha$ and $\beta$ are two positive constants. This model is
consistent with the cosmological and solar system tests
\cite{Nojiri}. It was also shown that the model (\ref{fR})
 can predict the unification of the early time
inflation and late time cosmic acceleration \cite{Noj4}.

Note that to investigate the GSL (\ref{S Tot}) for model (\ref{fR})
we need to have the scale factor $a(t)$. The scale factor can be
obtained from the first Friedmann equation (\ref{ro}). But before it
one needs to determine the evolution of the matter from the energy
equation (\ref{ec}). To do this if one takes the barotropic matter
$p_{\rm m}=\omega_{\rm m}\rho_{\rm m}$ with constant equation of
state (EoS) parameter $\omega_{\rm m}\geq0$, then solving the energy
equation (\ref{ec}) gives
\begin{equation}
\rho_{\rm m}=\rho_{\rm m_0}a^{-3(1+\omega_{\rm m})}.\label{ecm}
\end{equation}
Substituting Eqs. (\ref{fR}) and (\ref{ecm}) into (\ref{ro}) gives a
complicated nonlinear differential equation for the scale factor
that cannot be solved analytically. To avoid of this problem in what
follows we investigate the validity of the GSL (\ref{S Tot}) for
model (\ref{fR}) during the early inflation (large curvature) and
late acceleration (small curvature) eras. In the intermediate epoch
from strong to low curvature $R$, the universe subsequently enters
in the radiation and then matter dominated eras. In these two
epochs, the last two terms in Eq. (\ref{fR}) have the same orders of
magnitude as the first term. Therefore, model (\ref{fR}) recovers
the Einstein gravity in which the GSL is satisfied (see Eq.
\ref{GSL-R}). In what follows we further assume the universe to be
spatially flat, i.e. $K=0$, which is compatible with the
observations \cite{Riess}.
\subsection{The early inflation epoch}
During the early inflationary phase of the universe, the scalar
curvature $R$ is large and model (\ref{fR}) behaves like
\begin{equation}
f(R)\simeq\beta R^{2}.\label{fRinf}
\end{equation}
Here, we first consider the case of pure $f(R)$-gravity in which the
contribution of matter is neglected. In this case, inserting Eq.
(\ref{fRinf}) into (\ref{ro}) gives the de Sitter scale factor
\begin{equation}
a(t)\propto e^{Ht},~~~H={\rm constant}.\label{ainf1}
\end{equation}
Also the deceleration parameter is obtained as
\begin{equation}
q=-1-\frac{\dot{H}}{H^2}=-1.\label{q}
\end{equation}
Using Eqs. (\ref{fRinf}) and (\ref{ainf1}) the GSL (\ref{S Tot}) for
$K=0$ gives
\begin{equation}
T_{\rm A}\dot{S}_{\rm tot}=0,
\end{equation}
which corresponds to a reversible adiabatic expansion of the
universe.

Now we investigate the more realistic case in which the
$f(R)$-gravity to be minimally coupled with matter. Using Eqs.
(\ref{ecm}) and (\ref{fRinf}), solving Eq. (\ref{ro}) leads to the
scale factor
\begin{equation}
a(t) \propto t^{\frac{4}{3(1+\omega_{\rm m})}}.
\end{equation}
The corresponding Hubble and deceleration parameters are
\begin{equation}
H=\frac{4}{3(1+\omega_{\rm m})t},\label{Hinf}
\end{equation}
\begin{equation}
q=-1+\frac{3(1+\omega_{\rm m})}{4}.\label{qinf}
\end{equation}
Here, due to having an accelerating universe, i.e. $q<0$, from Eq.
(\ref{qinf}) we need to have $0\leq\omega_{\rm m}<1/3$.

Using Eqs. (\ref{fRinf}) and (\ref{Hinf}), the GSL (\ref{S Tot}) for
$K=0$ yields
\begin{equation}
T_{\rm A}\dot{S}_{\rm tot}=\frac{9\beta(1-3\omega_{\rm
m})(5-3\omega_{\rm m})}{4Gt^{2}},
\end{equation}
which shows that the GSL in the early inflation era is satisfied
when $\omega_{\rm m}\geq5/3$ or $0\leq\omega_{\rm m}\leq 1/3$.
However, the condition $\omega_{\rm m}\geq5/3$ should be avoided
because it does not yield an accelerated inflationary phase of the
universe (see Eq. \ref{qinf}). Note that $\omega_{\rm m}=0$ and 1/3
correspond to the pressureless dust matter (or DM) and radiation,
respectively.
\subsection{The late acceleration era}

In the late time epoch, the scalar curvature R is small. Hence, the
only contribution that can reproduce the late time acceleration in
model (\ref{fR}) is
\begin{equation}
f(R)\simeq -\alpha R^{-1}.\label{fRlate}
\end{equation}
For the pure $f(R)$-gravity case, i.e. ${\rm \rho_{\rm m}}\ll{\rm
\rho_{\rm R}}$, solving Eq. (\ref{ro}) yields
\begin{equation}
a(t) \propto t^{2},
\end{equation}
and the Hubble parameter is obtained as
\begin{equation}
H=\frac{2}{t}.\label{Hlate1}
\end{equation}
Here, the deceleration parameter is $q=-1/2$. Using Eqs.
(\ref{fRlate}) and (\ref{Hlate1}), the GSL (\ref{S Tot}) for $K=0$
leads to
\begin{equation}
T_{\rm A}\dot{S}_{\rm tot}=\frac{\alpha~t^{4}}{1152~G}>0,
\end{equation}
which shows that the GSL holds for the case of pure $f(R)$-gravity.\\

For the case of $f(R)$-gravity coupled with matter, inserting Eqs.
(\ref{ecm}) and (\ref{fRlate}) into (\ref{ro}) gives the scale
factor
\begin{equation}
a(t) \propto t^{\frac{-2}{3(1+\omega_{\rm m})}},\label{alate}
\end{equation}
which shows that the universe is shrinking in the presence of
ordinary matter ($\omega_{\rm m}\geq0$). But if we change the arrow
of time by $t\rightarrow t_{\rm s}-t$, the expansion occurs with the
inverse power law and at $t=t_{\rm s}$, the size of the universe
diverges. This modification yields the Hubble and deceleration
parameters as
\begin{equation}
H=\frac{2}{3(1+\omega_{\rm m})(t_{\rm s}-t)},~~~t\leq t_{\rm
s},\label{Hlate}
\end{equation}
\begin{equation}
q=-1-\frac{3(1+\omega_{\rm m})}{2}.\label{qlate}
\end{equation}
The above relation describes an expanding super-accelerating
($q<-1$) universe filled with ordinary matter ($\omega_{\rm
m}\geq0$).

Using Eqs. (\ref{fRlate}) and (\ref{Hlate}) the GSL (\ref{S Tot})
for $K=0$ yields
\begin{equation}
T_{\rm A}\dot{S}_{\rm tot}=\frac{243\alpha(1+\omega_{\rm
m})^{6}(11+6\omega_{\rm m})(t_{\rm s}-t)^4}{128G(7+3\omega_{\rm
m})^{2}},
\end{equation}
which clarifies that the GSL is always satisfied for the case of
$f(R)$-gravity coupled with ordinary matter ($\omega_{\rm m}\geq0$).
\section{Conclusions}

Here, we studied the GSL of gravitational thermodynamics in the
framework of $f(R)$-gravity. Among other approaches related with a
variety of DE models, a very promising approach to DE is related
with the modified theories of gravity known as $f(R)$-gravity, in
which DE emerges from the modification of geometry. Modified gravity
gives a natural unification of the early time inflation and late
time acceleration thanks to different role of gravitational terms
relevant at large and at small curvature and may naturally describe
the transition from deceleration to acceleration in the cosmological
dynamics. We considered a FRW universe containing only the ordinary
matter with positive constant EoS parameter $\omega_{\rm m}$. The
boundary of the universe was assumed to be enclosed by the dynamical
apparent horizon with the Hawking temperature. We examined the
validity of the GSL for a viable model as $f(R)=R-\alpha/R+\beta
R^{2}$ in the two regimes containing the early inflation and late
acceleration eras. In the intermediate epoch containing the
radiation and matter dominated eras, the GSL is always satisfied. We
concluded that in the early time, the GSL for the pure
$f(R)$-gravity case (with no matter) behaves like a reversible
adiabatic expansion of the de Sitter universe. Furthermore, for the
case of $f(R)$-gravity minimally coupled with matter the GSL is
satisfied only for $0\leq\omega_{\rm m}\leq 1/3$. In the late time,
for the case of pure $f(R)$-gravity the GSL holds. Also for the case
of $f(R)$-gravity coupled with ordinary matter the GSL is always
satisfied.
\section*{Acknowledgements}
The work of K. Karami has been supported financially by Research
Institute for Astronomy and Astrophysics of Maragha (RIAAM) under
research project No. 1/2342.

\end{document}